\documentclass[3p,times]{elsarticle}

\usepackage{ecrc}

\usepackage{epstopdf}
\usepackage{xspace}

\newcommand{\jpsi}{\ensuremath{{J/\psi}}\xspace}

\newcommand{\pT}{\ensuremath{p_{\rm T}}\xspace}

\newcommand{\mT}{\ensuremath{m_{\rm T}}\xspace}

\newcommand{\mee}{\ensuremath{m_{ee}}\xspace} 
\newcommand{\s}{\ensuremath{\sqrt{s}}\xspace}

\newcommand{\snn}{\ensuremath{\sqrt{s_{\rm NN}}}\xspace}

\volume{00}

\firstpage{1}

\journalname{Nuclear Physics A}

\runauth{M.~K.~K\"{o}hler et al.}

\jid{nupha}

\jnltitlelogo{Nuclear Physics A}

\usepackage{graphicx}
\usepackage{amsmath,amssymb}

\usepackage{lineno}

\begin{document}

\begin{frontmatter}



\title{Dielectron measurements in pp, p--Pb and Pb--Pb collisions with ALICE at the LHC}

\author{Markus K. K\"{o}hler\\ (for the ALICE\fnref{col1} Collaboration)}
\fntext[col1] {A list of members of the ALICE Collaboration and acknowledgements can be found at the end of this issue.}
\address{Research Division and ExtreMe Matter Institute EMMI,\\ GSI Helmholtzzentrum f\"{u}r Schwerionenforschung, Germany}




\begin{abstract}
Electromagnetic probes are excellent messengers from the hot and dense medium created in high-energy heavy-ion collisions. Since
leptons do not interact strongly, their spectra reflect the entire space-time evolution of the collision. The surrounding medium can lead to
modifications of the dielectron production with respect to the vacuum rate. To quantify modifications in heavy-ion collisions, measurements
in pp collisions serve as a reference, while the analysis of p--A collisions allows for the disentanglement of cold nuclear matter effects
from those of the hot and dense medium. 

In this proceedings, dielectron measurements with the ALICE central barrel detectors are presented. The invariant mass distributions in the
range $ 0<m_{ee}<3 $~GeV/$ c^{2} $ are compared to the expected yields from hadronic sources for pp collisions at $ \sqrt{s}=7 $~TeV, and
for p--Pb collisions at $ \sqrt{s_{\rm NN}}=5.02 $~TeV. The cross section of direct photons measured via virtual photons in pp
collisions is compared to predictions from NLO pQCD calculations as a function of the transverse momentum. The status of the analysis of
Pb--Pb collisions at $ \sqrt{s_{\rm NN}}=2.76 $~TeV is presented.

\end{abstract}


\end{frontmatter}



\section{Introduction}
\label{intro}
Dielectrons were proposed several decades ago~\cite{Shuryak1978} to be an important source of information from the hot and
dense medium, which can be created in heavy-ion collisions. Since dielectrons are emitted throughout the collision process and do not
interact via the strong interaction, they are ideal probes for all stages of the collision. Moreover, the measurement of virtual
photons, i.e. photons which convert internally into dileptons, allows to reduce systematic uncertainties significantly compared to the
measurement of real photons, since the main sources of the background, photons and dielectrons from $\pi^0$ decays, can be rejected at
finite mass.

However, to access information on the medium in heavy-ion collisions, the dielectron production in the vacuum and possible cold nuclear
matter effects need to be evaluated. Therefore, it is necessary to have reference measurements from proton-proton (pp) and proton-nucleus
(p--A) collisions. 

LHC provided during Run~1 three different collision systems, i.e. pp, p--Pb and Pb--Pb. In this proceedings, preliminary results from pp
collisions at $\s=7$~TeV on the dielectron invariant mass continuum and direct photons measured via virtual photons are summarized. The
dielectron continuum as a function of invariant mass and dielectron transverse momentum is compared to the expected hadronic sources of
dielectrons in p--Pb collisions at $\snn = 5.02$~TeV. In addition, the status of the analysis for Pb--Pb collisions at $\snn = 2.76$~TeV
will be discussed.
\section{Data analysis}
\label{analysis}
ALICE has capabilities for particle identification in the low transverse momentum ($\pT$) regime that are unique at the LHC. Electrons with
the transverse momentum $\pT^e > 0.2$~GeV/$c$ are identified by combined energy loss information from the Time Projection Chamber (TPC)
and, in the case of p--Pb and Pb--Pb, the outermost four layers of the Inner Tracking System (ITS). Additionally, the Time-Of-Flight
detector (TOF) is used in the range $ 0.4 < \pT^e < 5.0$~GeV/$c$ to reject kaons and protons. The remaining hadron contamination is at most
$1$~\% in pp collisions and up to $10$~\% in Pb--Pb collisions.

When measuring unlike-sign dielectron pairs, $N_{\rm US}$, one of the main challenges is the estimation of the combinatorial background,
which arises from random dielectron combinations and is superimposed on the physics signal. The signal-over-background ratio is in
the order of $10^{-2}$ for pp and p--Pb collisions and about a factor $10$ lower in central Pb--Pb collisions for $\mee \approx
0.5$~GeV/$c^2$. The combinatorial background is measured by the same-event like-sign method. This method holds under the assumption that
the physics signal consists only of unlike-sign pairs. The like-sign spectra are normalized via $N_{\rm LS} = 2 \cdot
R \cdot \sqrt{N_{++}N_{--}}$, where $R$ is a correction factor for the difference between the acceptance of unlike-sign pairs and like-sign
pairs and $N_{++}$ and $N_{--}$ are positive and negative like-sign dielectron pairs, respectively. The acceptance correction is calculated
as $R = B_{+-}/(2 \cdot \sqrt{B_{++}B_{--}})$, where $B$ indicates mixed event distributions. $R$ depends on the minimum single
electron $\pT^e$ and is consistent with unity within its statistical uncertainties for the pp and the p--Pb analysis for $\pT^e >
0.2$~GeV/$c$. For $\pT^e > 0.4$~GeV/$c$ in Pb--Pb collisions, the deviation from unity is of the order of $5$~\% for $\mee < 0.1$~GeV/$c^2$
and approaches unity for increasing mass. The raw signal is calculated as $ S = N_{\rm US} - N_{\rm LS}$.

The data are corrected for detector and reconstruction efficiency via Monte-Carlo (MC) simulations. Single electron efficiencies are
calculated as a function of $(\pT,\eta,\phi)$. Every electron is weighted with its efficiency in a dielectron generator with realistic
electron and dielectron kinematics.

\begin{figure}[t]
\begin{minipage}{14pc}
\includegraphics[scale = 0.38]{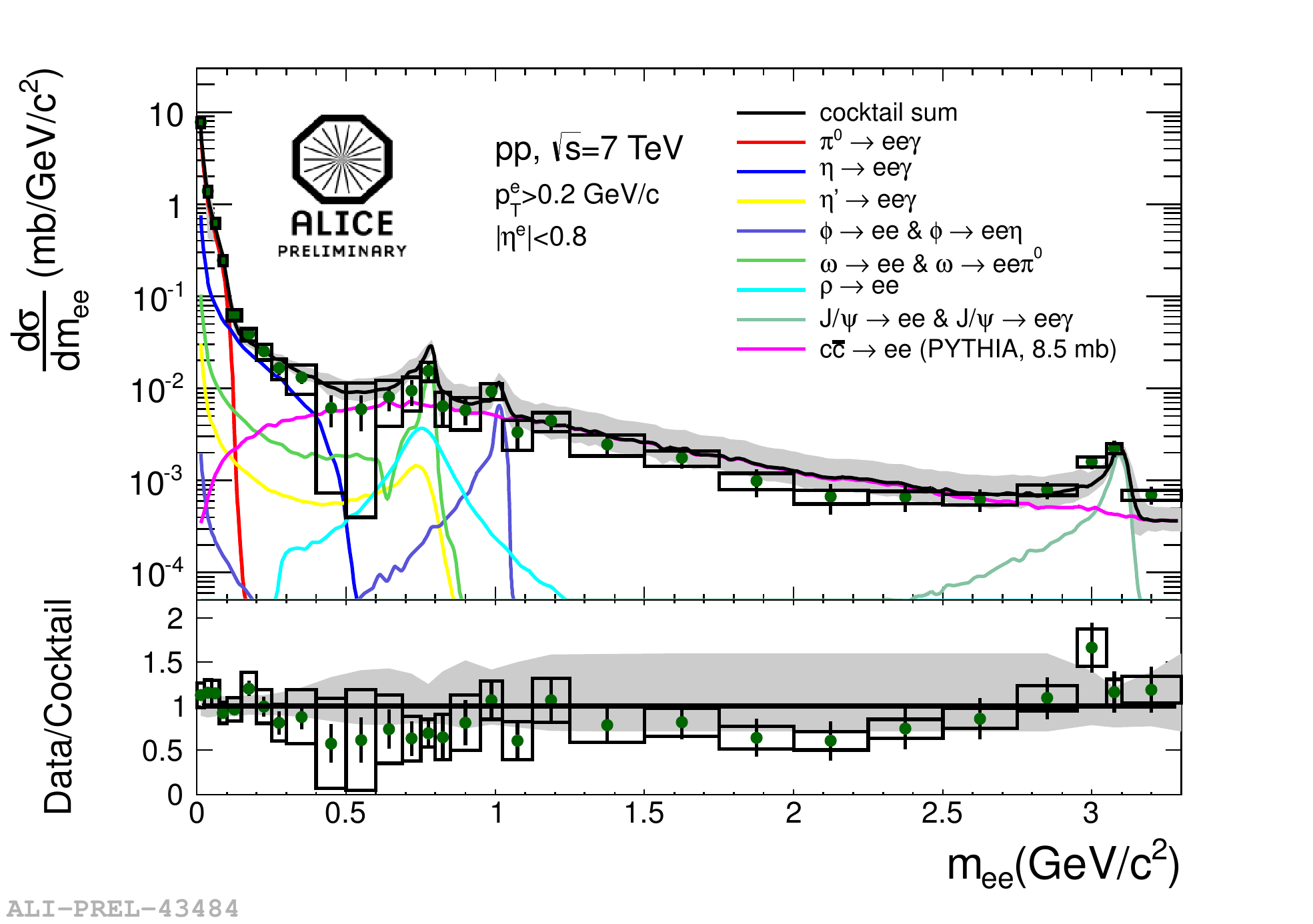}
\end{minipage}\hspace{4pc}%
\begin{minipage}{14pc}
\includegraphics[scale = 0.38]{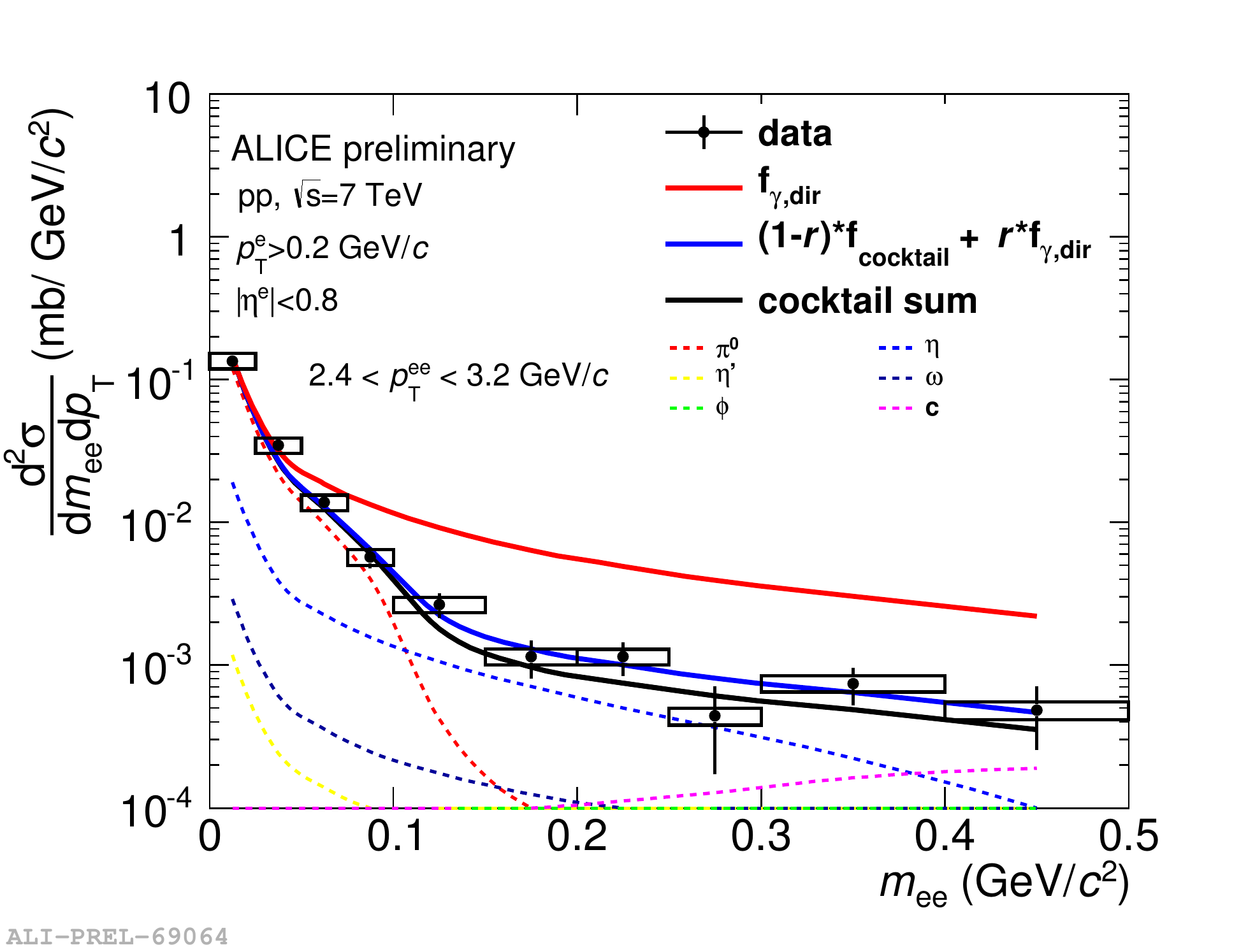}
\end{minipage}
\begin{minipage}{14pc}
\includegraphics[scale = 0.38]{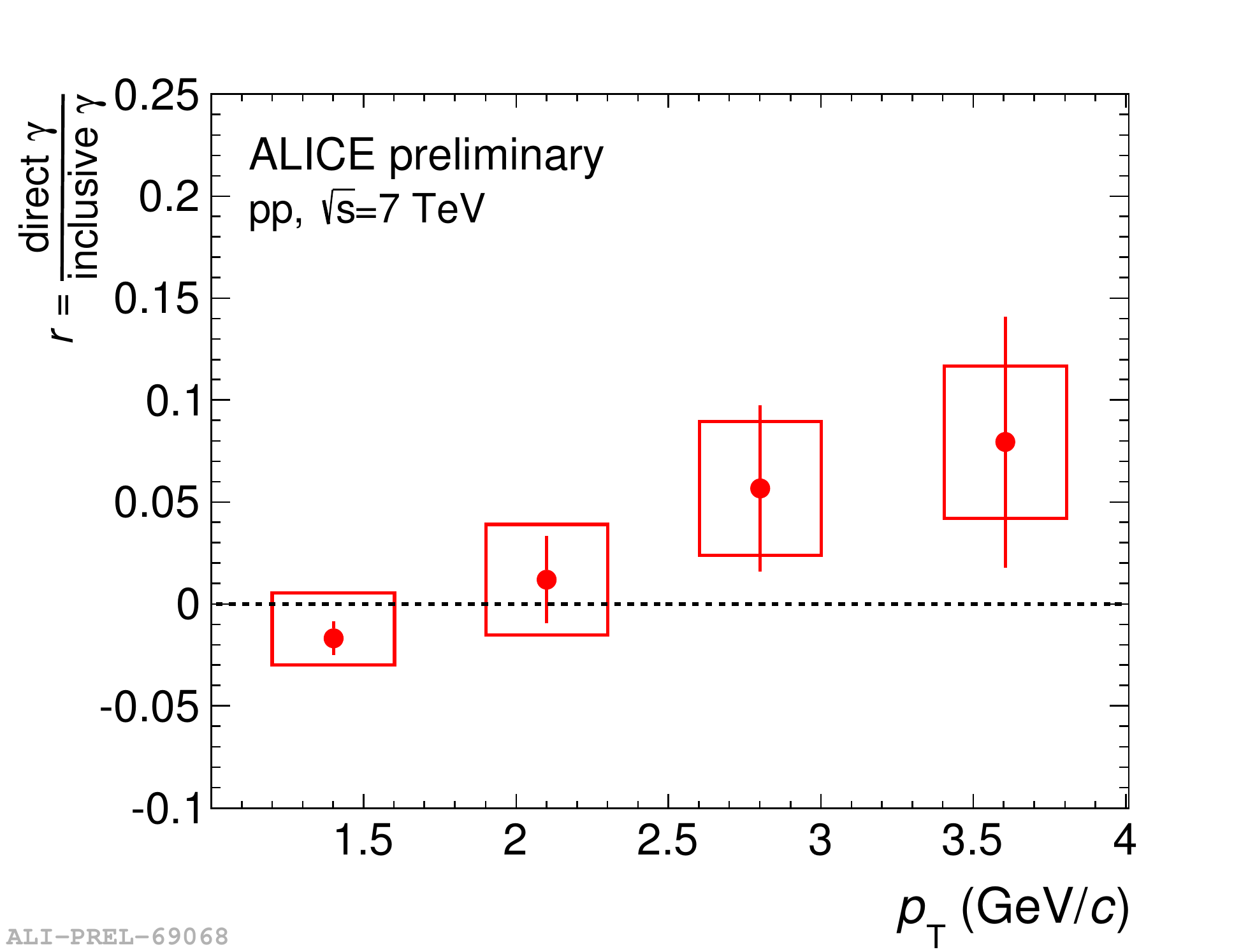}
\end{minipage}\hspace{4pc}%
\begin{minipage}{14pc}
\includegraphics[scale = 0.35]{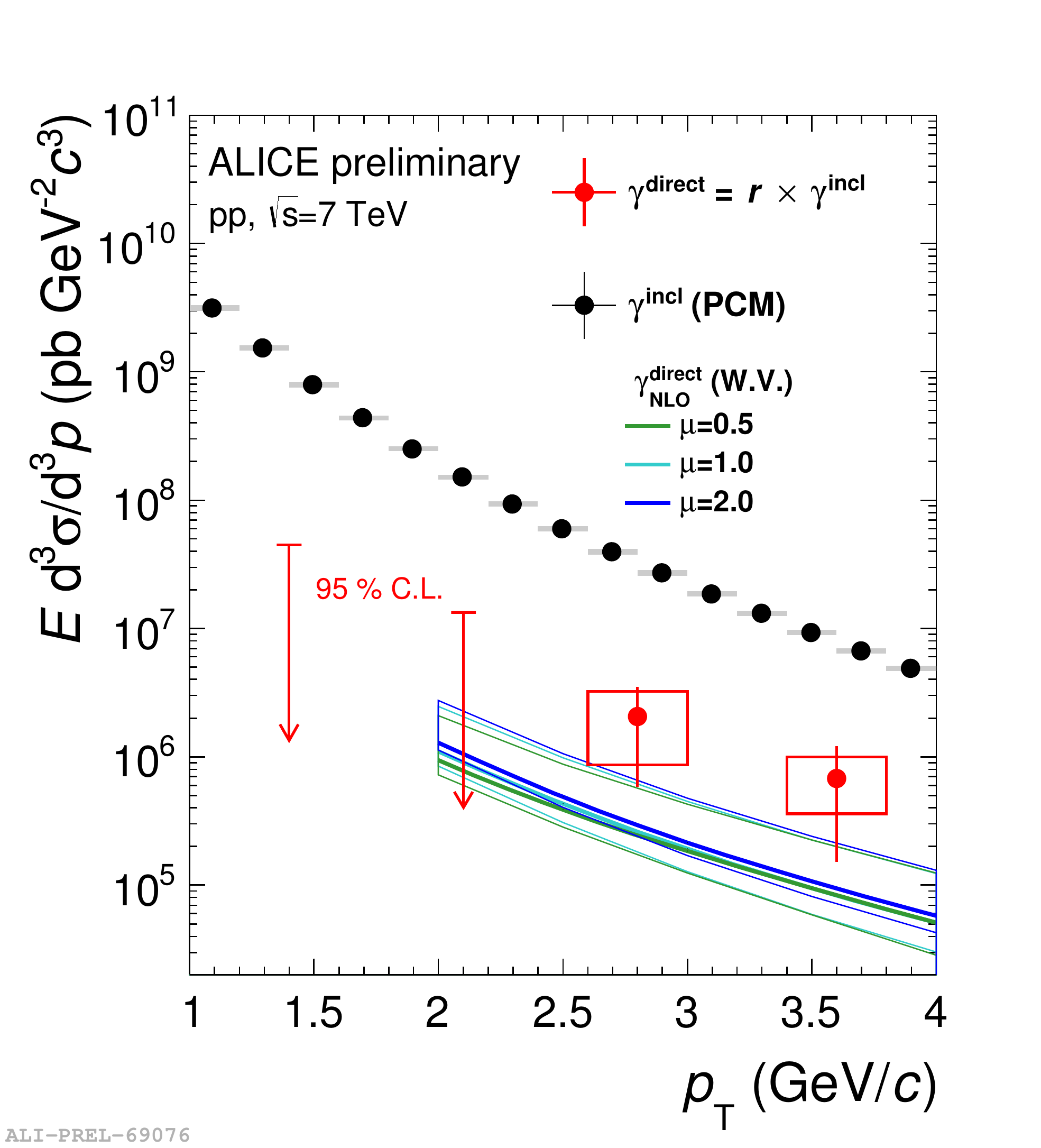}
\end{minipage} 
\caption{Upper left: Dielectron invariant mass distribution together with the cocktail calculations for pp collisions at $\s = 7$~TeV. Upper
right: Result for two component fit of the \mee distribution. Lower left: Fit parameter $r$ as a function of photon \pT. Lower right:
Direct photon cross section as a function of transverse momentum together with pQCD NLO
calculations. At low \pT only upper limits (95~\% confidence level) can be determined, as indicated by the arrows. The inclusive photon
cross section measured via photon conversion method (PCM) is also shown.}\label{fig:pp_plots}
\end{figure}
The expected hadronic sources of dielectrons at the moment of freeze-out, the so-called hadronic cocktail, are calculated based on
measured differential cross sections for $\pi^0,\eta,\phi$ and $\jpsi$ in pp collisions, and on the charged pion spectrum in p--Pb
collisions. The mass shape of resonances is based on~\cite{Gounaris} and the Dalitz pair mass distributions are following~\cite{KrollWada}.
To estimate the expected yield of correlated dielectrons from semi-leptonic decays of open heavy-flavor mesons, the PYTHIA
event generator~\cite{pythia}, tuned to NLO calculations~\cite{mangano}, is used. The pair distributions are normalized to the cross
sections measured in pp collisions and scaled by the number of binary collisions for p--Pb collisions. See~\cite{ALICE_cocktail} for a
compilation of references. Contributions from hadrons, which have not been measured, are estimated by $\mT$ scaling of the $\pi^0$ cross
section. 

\section{Results}
\label{results}
The preliminary results for pp collisions at $\s = 7$~TeV are shown in Figure~\ref{fig:pp_plots}. In the upper left panel, the dielectron
data are compared to the cocktail as a function of invariant mass for integrated pair \pT. The hadronic cocktail is consistent with the
dielectron data. Virtual photon production is studied in pp collisions. Virtual photons convert internally into dielectrons. The relation
between the dielectron invariant mass distribution and the virtual photon yield is given for $\pT^{ee} \gg \mee$ by the Kroll-Wada
equation~\cite{KrollWada}.

In the upper right panel of Figure~\ref{fig:pp_plots}, pp data are compared to the cocktail for $2.4 < \pT^{ee} < 3.2$~GeV/$c$.
The different sources are indicated as dashed lines. The function $f_{\rm comb} = (1-r)f_{c} + rf_{\rm \gamma,dir}$ is fitted to the data in
the region $0.1 < \mee < 0.4$~GeV/$c^2$, where $f_{c}$ is the cocktail contribution, $f_{\rm \gamma,dir}$ is the photon input from the
Kroll-Wada equation and $r$ is the only fitting parameter. $r$ reflects the ratio of direct over inclusive photons. The result for $r$ as a
function of the photon \pT is shown in the lower left panel of Figure~\ref{fig:pp_plots}. Under the assumption that the ratio of
direct over inclusive photons is the same for real and virtual photons, the direct photon cross section can be calculated by $\gamma_{\rm
dir} = r \times \gamma_{\rm incl}$, where $\gamma_{\rm dir}$ and $\gamma_{\rm incl}$ are the direct and inclusive photon yields. The
inclusive photon cross section has been measured via photon conversions, see e.g.~\cite{Wilde}. In the lower right panel of
Figure~\ref{fig:pp_plots}, the direct photon cross section is shown as a function of the photon \pT. NLO pQCD
calculations~\cite{Vogelsang} are consistent with the data.

\begin{figure}[t]
\begin{minipage}{14pc}
\includegraphics[scale = 0.34]{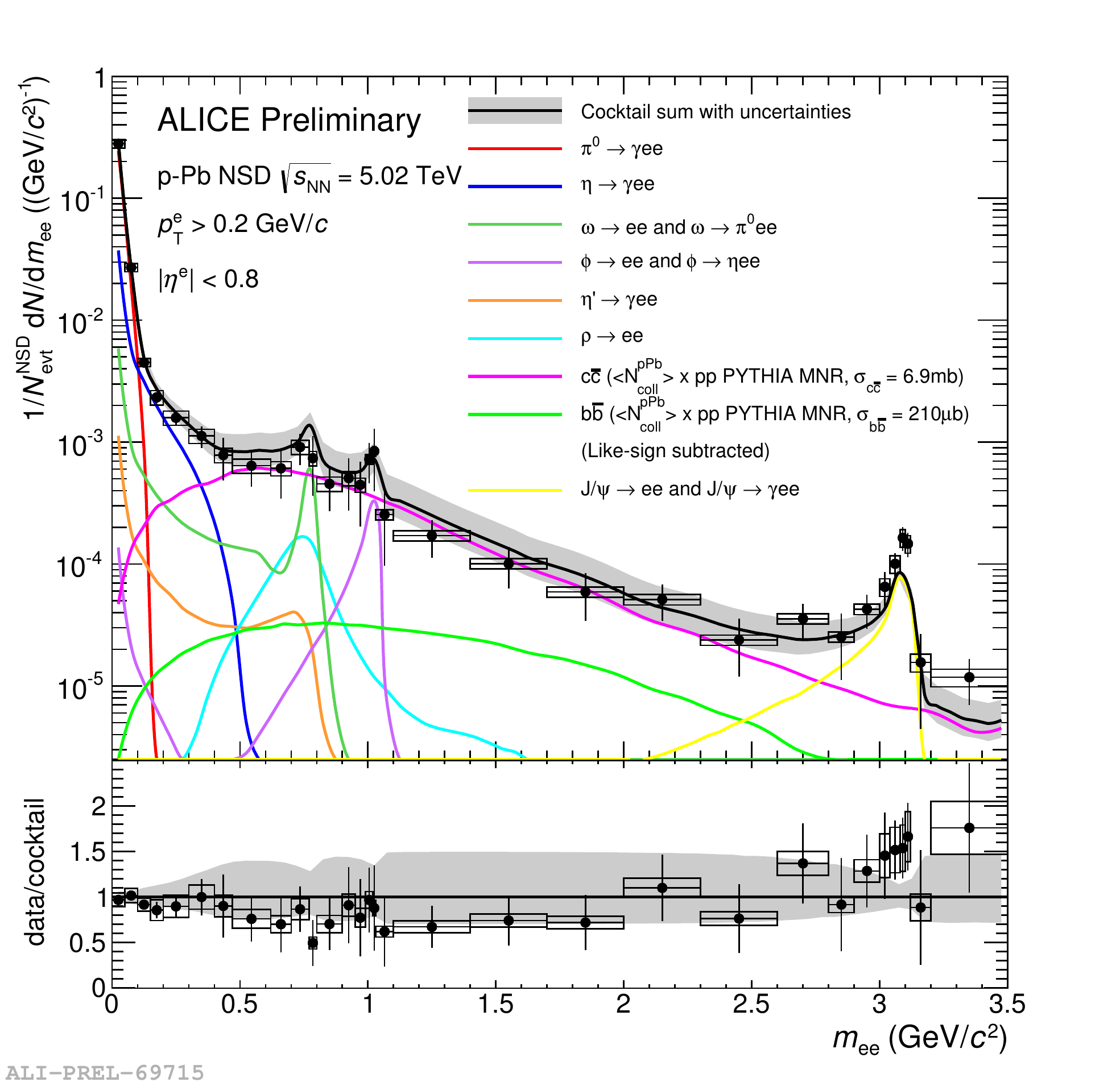}
\end{minipage}\hspace{4pc}%
\begin{minipage}{14pc}
\includegraphics[scale = 0.34]{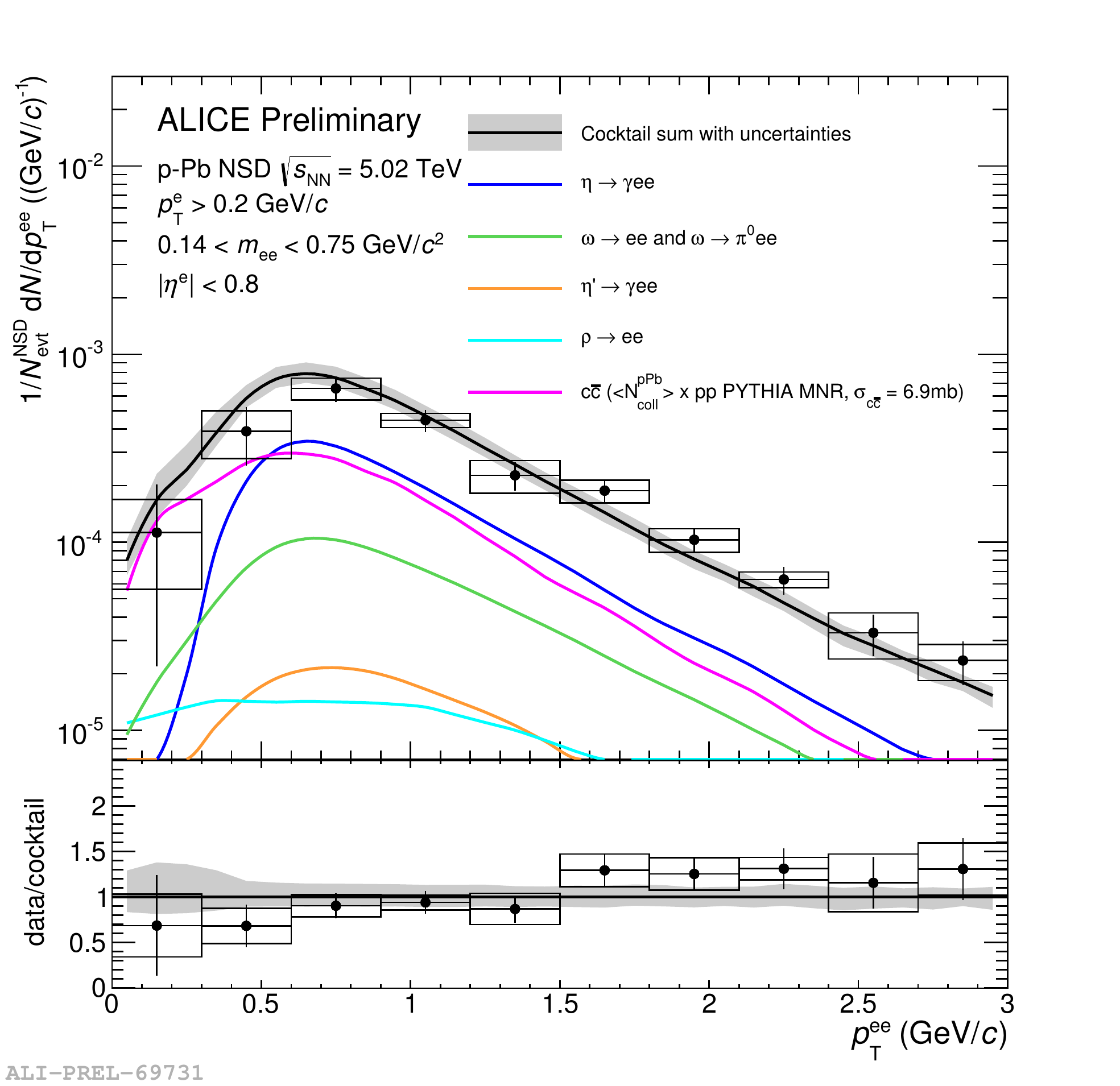}
\end{minipage}
\begin{minipage}{14pc}
\includegraphics[scale = 0.34]{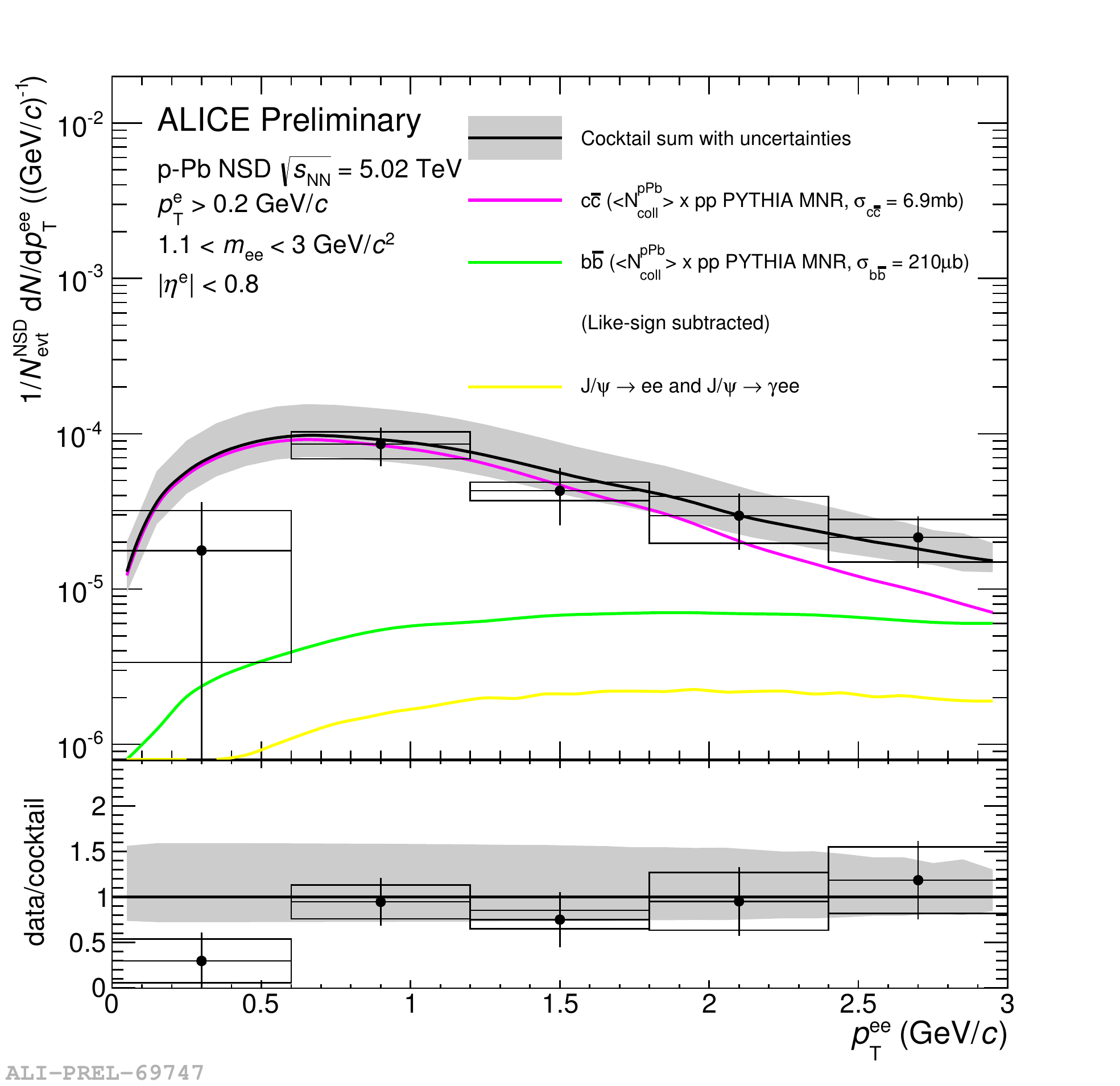}
\end{minipage}\hspace{4pc}%
\begin{minipage}{14pc}
\includegraphics[scale = 0.34]{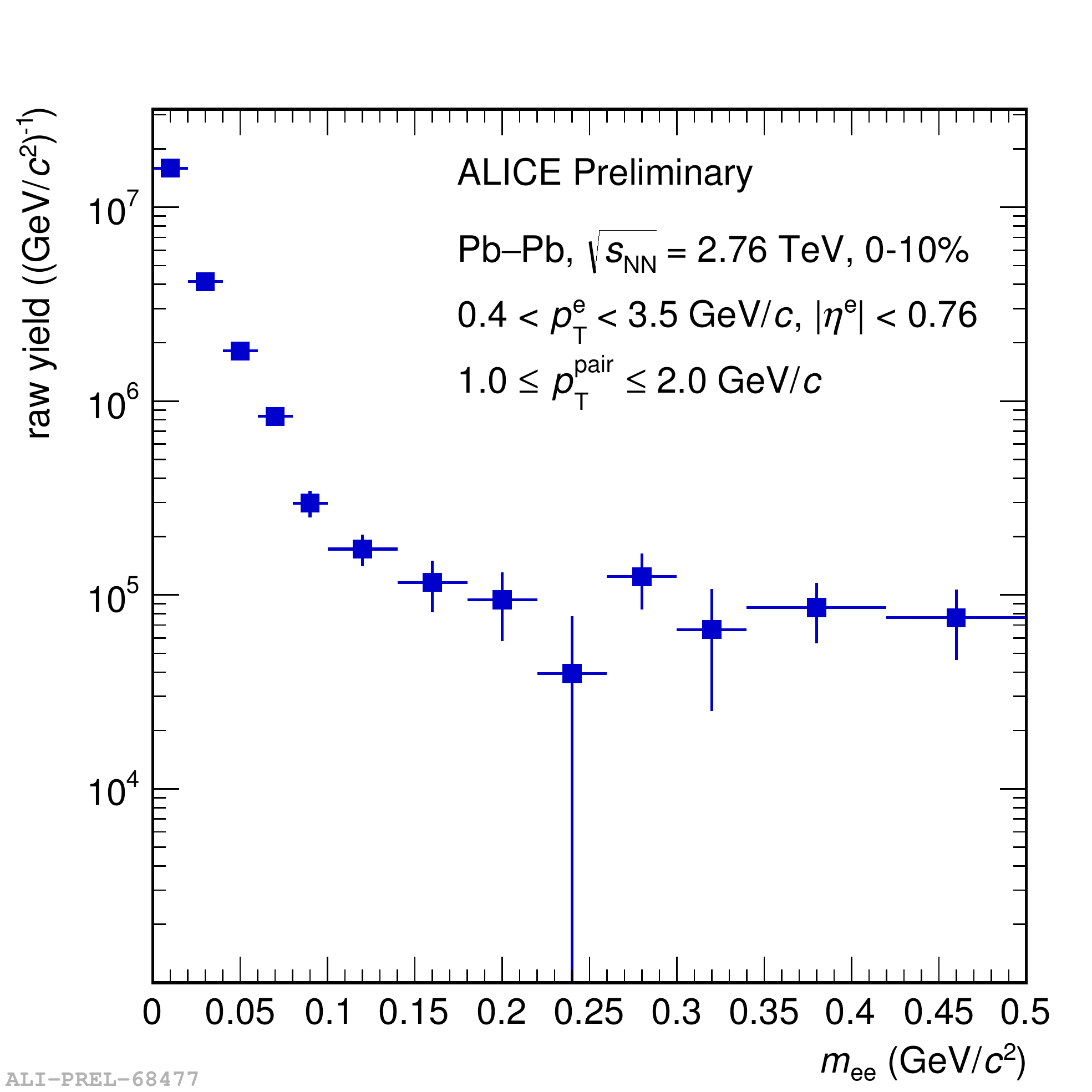}
\end{minipage} 
\caption{The dielectron mass distribution is compared to the cocktail calculations for p--Pb collisions at $\snn = 5.02$~TeV as a
function of invariant mass (upper left panel) and as a function of pair transverse momentum for the mass intervals $0.14 < \mee <
0.75$~GeV/$c^2$ (upper right panel) and $1.1 < \mee < 3.0$~GeV/$c^2$ (lower left panel). In the lower right panel, the raw dielectron yield
is shown for Pb--Pb collisions at $\snn = 2.76$~TeV.}\label{fig:pPb_plots}
\end{figure}

In the upper left panel of Figure~\ref{fig:pPb_plots}, the dielectron invariant mass spectrum is compared to the cocktail for p--Pb
collisions at $\snn = 5.02$~TeV. The cocktail is in good agreement with the data. In the upper right and lower left panels, data are
compared to the cocktail distributions as a function of dielectron transverse momentum for $0.14 < \mee < 0.75$~GeV/$c^2$ and $1.1 < \mee <
3.0$~GeV/$c^2$, respectively. The data are well described by the cocktail. These two mass regions are of special interest. The mass
region $0.14 < \mee < 0.75$~GeV/$c^2$ is sensitive to hot hadronic medium effects. The region $1.1 < \mee < 3.0$~GeV/$c^2$ is dominated by
semi-leptonic decays of heavy-flavour mesons. In this mass region, heavy quark pair correlations can be studied.

In the lower right panel of Figure~\ref{fig:pPb_plots}, the raw yield as a function of invariant mass is shown for Pb--Pb collisions in
$0-10$~\% centrality at $\snn = 2.76$~TeV. Further analysis of this spectrum will allow the study of the virtual photon yield and for
the exploration of a possible low-mass enhancement in Pb--Pb collisions.

\section{Summary and outlook}
\label{conclusions}
The dielectron invariant mass spectrum measured in pp collisions at $\s  = 7$ TeV is consistent with the expectation from hadronic sources.
The same is observed for the invariant mass and transverse momentum distributions of dielectrons in p--Pb collisions at 5.02 TeV. The direct
photon yield extracted from dielectron data in pp collisions at 7 TeV is consistent with NLO pQCD calculations. The study of dielectron
production in Pb--Pb collisions at $\snn = 2.76$~TeV is ongoing.  \\
At the end of Run~$2$, statistical uncertainties will be reduced significantly. After the second long shutdown at the LHC, which is expected
to end in $2019$, ALICE will run with upgraded detector components~\cite{ALICE_upgrade}. The upgrade of the ITS will allow high precision
vertexing to measure and reject dielectrons from correlated heavy flavour decays and the continuous read-out of the TPC will allow to take
full advantage of the high luminosity at the upgraded LHC. Hence, detailed studies of the dielectrons will become feasible in Pb--Pb
collisions.




\begin{thebibliography}{00}


\bibitem{Shuryak1978} E.~V.~Shuryak, Phys. Lett. B78 (1978) 150
\bibitem{Gounaris} G. Gounaris and J. Sakurai, Phys. Rev. Lett. 21 (1968) 244
\bibitem{KrollWada} N. Kroll and W. Wada, Phys. Rev. 98 (1955) 1355
\bibitem{pythia} T. Sjostrand, S. Mrenna and P. Skands, JHEP 05 (2006) 026
\bibitem{mangano} M. Mangano, P. Nason, and G. Ridolfi, Nucl. Phys. B 373 (1992) 295
\bibitem{ALICE_cocktail} ALICE Collaboration, Phys. Lett. B717 (2012) 162 ; Eur. Phys. J. C72(2012)2183 ; Phys. Lett. B704 (2011) 442 ;
JHEP 1207 (2012) 116
\bibitem{Vogelsang} W.~Vogelsang, private communication
\bibitem{Wilde} M.~Wilde (for the ALICE Collaboration), arXiv:1210.5958 [hep-ex] (2012)
\bibitem{ALICE_upgrade} ALICE Collaboration, J. Phys. G 41 (2014) 087001 ; CERN-LHCC-2012-012 ; LHCC-I-022 (2012)

\end{thebibliography}



\end{document}